\documentclass[12pt]{iopart}

\def\>{\rangle}
\def\<{\langle}
\def\ket#1{|#1\>}
\def\bra#1{\<#1|}
\def\braket#1#2{\< #1 | #2 \>}
\def\half{\frac{1}{2}}
\def\quar{\frac{1}{4}}

\usepackage{psfig}

\begin{document}

\title{Many-body symbolic dynamics of a classical oscillator chain}

\author{Marko \v Znidari\v c \footnote{znidaricm@fiz.uni-lj.si} and
Toma\v z Prosen \footnote{prosen@fiz.uni-lj.si}}

\address{Physics Department, Faculty of Mathematics and Physics, 
University of Ljubljana, Jadranska 19,1111 Ljubljana, Slovenia}

\begin{abstract}
We study a certain type of the celebrated Fermi-Pasta-Ulam particle
chain, namely the inverted FPU model, where the inter-particle 
potential has a form of a quartic double well.
Numerical evidence is given in support of a simple symbolic description
of dynamics (in the regime of sufficiently high potential barrier between
the wells) in terms of an (approximate) Markov process. The corresponding 
transition matrix is formally identical to a ferromagnetic Heisenberg quantum
spin-$1/2$ chain with long range coupling, whose diagonalization yields accurate 
estimates for a class of time correlation functions of the model.
\end{abstract}

\pacs{05.45.-a}

\maketitle

\section{Introduction}

In a historical numerical experiment Fermi Pasta and Ulam \cite{FPU} 
studied the following non-linear model of a 1D crystal described
by the Hamiltonian
\begin{equation}
H=\sum_{i=1}^N\left( \half m \dot{x}_i^2 + V(x_i-x_{i-1}-l)\right),\quad
V(x) = \half m \omega^2 x^2 + \quar k x^4.
\label{dimH}
\end{equation}
Contrary to initial expectations of Fermi {\em et al} the 
so-called FPU model (\ref{dimH}) behaved in a strong contrast 
to the ergodic theorem of statistical mechanics, even in quite 
strongly non-linear regime when one would expect fast relaxation 
to statistical canonical equilibrium and equipartition of energy 
from an arbitrary initial state. Instead, FPU model triggered the 
discovery of non-linear normal modes, the so-called {\em solitons}, 
and indirectly, the whole field of (computational) non-linear dynamics. 
Later, FPU chain has been studied in the spirit of KAM theory 
\cite{KAM} and various estimates have been given for the critical 
strength of the dimensionless non-linearity parameter 
$k l^2/m\omega^2$ required for the motion to become globally stochastic 
\cite{Chirikov,Shepelyansky}. 
\par
Recently, FPU-type models have been studied in the context
of energy transport and Fourier heat law in 1D chains of 
particles \cite{heatFPU,hatano,Huetal}. Quite unexpectedly, FPU-type models, 
namely Hamiltonians of the type (\ref{dimH}) with non-linear
inter-particle interaction $V(x)$ (and no {\em on-site} potential), 
turned out to be anomalous heat conductors, due to slow power law
decay of transport (velocity-velocity or current-current) 
time correlation functions. These results, namely that correlations
decay universally as $ C(t) \sim t^{-3/5} $, and correspondingly that Kubo
transport coefficient diverges as $\kappa(L)\sim L^{2/5}$ as a function
of the chain length $L$, have later been successfully
explained in terms of hydrodynamic arguments and mode-mode coupling
theory \cite{modecoupl}.
\par 
However, a perturbative-like approach such as mode-mode coupling
theory should break down when the inter-particle potential has 
{\em more than one} stable position. Indeed, Giardina {\em et al}
\cite{Giardina} have performed a series of numerical experiments on the
particle chain with an oscillating inter-particle potential $V(x) = 1-\cos(x)$, 
as well as with the potential $V(x)=-x^2/2+x^4/4$, and 
found normal heat conduction with a clean exponential decay of correlations.
The key mechanism, which we believe produces non-KAM-like 
(almost) hyperbolic motion of such a high dimensional Hamiltonian system, 
are the hyperbolic saddles over which pairs of particles flip from one
well to another. This motivated us to study in this paper some 
fundamental dynamical properties of the simplest version of such a 
hyperbolic chain with quartic double-well inter-particle potential $V(x)$.
We present here some intriguing numerical results which suggest
existence of a simple (but only approximate) Markov partition with very
simple many-body symbolic dynamics
of this particle chain. In addition, we find that the
transition matrix is formally generated by a Hamiltonian of a certain quantum
spin-1/2 chain. 
\par
Section \ref{SecIFPU} introduces the model. In section \ref{SecAMP}
we define (approximate Markov) partition of phase space and obtain numerical and 
analytical estimates for the volumes of various cells. Expression for the 
average time between subsequent jumps among the cells is also obtained. 
In section \ref{SecMarkov} we introduce Markovian description of our model with
the transition matrix and the flux matrix and numerically check the 
accuracy of the Markovian property in two independent ways. 
Section \ref{SecF} represents the core of the present paper. 
We numerically calculate the transition matrix and make a formal
correspondence between our model and a ferromagnetic Heisenberg quantum
spin-$1/2$ chain. Using this correspondence we derive various estimates 
and bounds for the decay rates of the time correlation
functions of a class of piece-wise constant functions. 
  
\section{Model: Inverted Fermi-Pasta-Ulam chain}
\label{SecIFPU}
 
With the choice for units of mass, time and length of $m$, $1/\omega$ 
and $l$, respectively, and the new canonical coordinates 
$q_i=x_i-il$ and $p_i=m\dot{x_i}$, 
Hamiltonian (\ref{dimH}) can be brought to a dimensionless form 
\begin{equation}
H=\sum_{i=1}^N{\frac{p_i^2}{2}+V(q_i-q_{i-1})}, \qquad V(x)=
-\alpha \frac{x^2}{2}+\frac{x^4}{4}+\frac{\alpha^2}{4},
\label{aFPU}
\end{equation}
where $\alpha=-{m \omega^2}/{k l^2}$ is the only dimensionless 
parameter left. In order for the potential minima to be at level 0 (for $\alpha>0$) we have added a constant term $\alpha^2/4$ to the potential $V(x)$. 
We can get all different orbits of the system described by the Hamilton 
function (\ref{aFPU}) just by varying one parameter $\alpha$. 
At first it would seem that the energy $E$ is also a free 
parameter.
But by fixing all the three basic units we 
have also provided a natural unit for the energy density 
$\varepsilon=E/N$. If we scale the dimensionless quantities 
by a factor $\xi$, namely $(q_i,p_i,\alpha) 
\to (\xi q_i,\xi^2 p_i,\xi^2 \alpha)$, the Hamiltonian is simply
multiplied by a factor $\xi^4$. Invariant parameter in this 
transformation is therefore $\lambda=\varepsilon/\alpha^2$. 
All different nonequivalent Hamiltonians can be obtained just by 
varying one parameter, namely $\lambda$, or equivalently $\alpha$ at fixed
$\varepsilon$. 
Energy density can therefore be fixed without 
any loss of generality. From now on $\varepsilon=E/N=1$ is assumed, 
as well as periodic boundary conditions $(q_{N+1},p_{N+1})\equiv(q_1,p_1)$.
Phase space point will be denoted by $\bi{x}=(\bi{q},\bi{p})$.
For negative values of the parameter 
$\alpha$ the system has a form of the standard $\beta$-FPU model (just
in a different parametrization), but in this paper we focus on the 
positive values of the parameter $\alpha$ in which case we call the system 
{\em Inverted FPU model} (IFPU). IFPU model is translationally invariant, therefore besides the 
energy $\varepsilon$ there is also a second (trivial) constant of 
motion, namely the total momentum $P=\sum_{i=1}^N{p_i}$. 
In all numerical calculations the total momentum has been 
fixed to $P=0$.
\begin{figure}[!ht]
\centerline{\psfig{figure=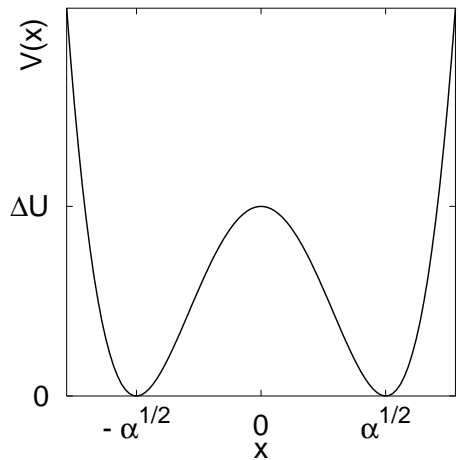,height=60mm,angle=-90}}
\caption{Potential $V(x)$ for the inverted FPU model (\ref{aFPU}), $\alpha>0$.}
\label{Slikapot}
\end{figure}
\par
For the numerical integration of equations of motion a fourth order 
symplectic algorithm of Ref. \cite{McLachlanAtela} has been used. This integrator 
has been checked to be the optimal choice (at least for the model studied here,
and at relative accuracy $\sim 10^{-5}-10^{-8}$)
by making a careful comparison with a number of other symplectic and 
Runge-Kutta methods.

\section{Phase space partition and statistical dynamics}
\label{SecAMP}

\begin{figure}[!ht]
\centerline{\psfig{figure=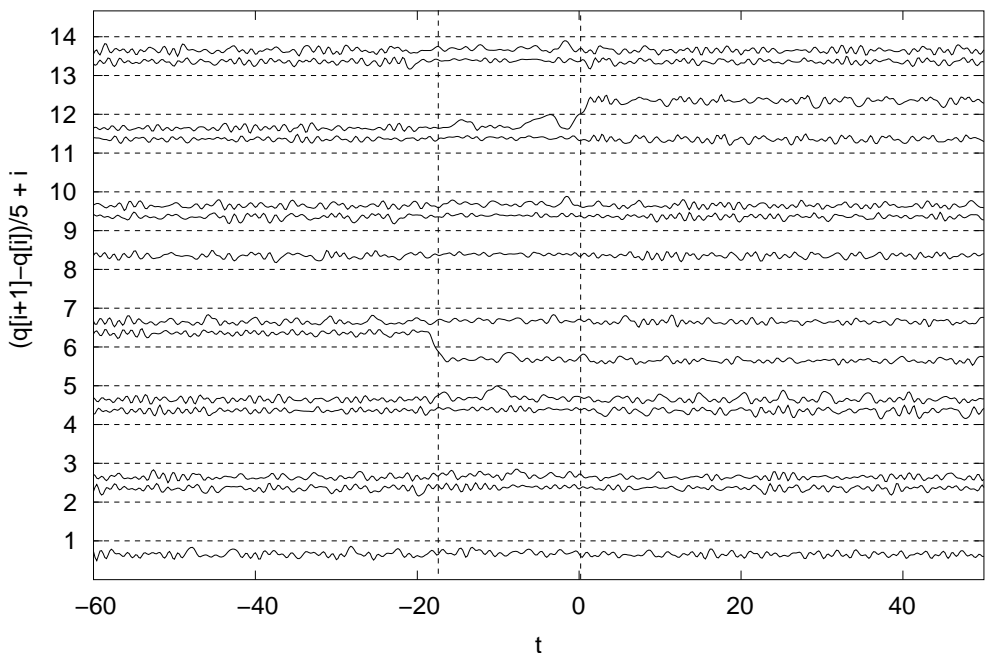,width=130mm,angle=-90}}
\caption{Time dependence of differences $(q_{i+1}(t)-q_i(t))$ for $N=14,\alpha=4.8$. For time $t<-20$ the 
system is in the order 0 state $S=01010110101010_2$, at $t\approx -20$ the system 
jumps to 
the temporary order 1 metastable state $S''=01010110001010_2$ and shortly thereafter at 
$t=0$ jumps into the new stable order 0 state $S'=01110110001010_2$. Such a transition from 
$S$ to $S'$ will be called a jump of length $d=6$. Dashed lines denote the position of the saddles $q_{i+1}=q_i$.}
\label{Slikaq}
\end{figure}

If the barrier $\Delta U = \alpha^2/4$ between the potential 
minima is sufficiently high the differences $(q_{i+1}(t)-q_i(t))$ will 
spend most of the time either around the left or the right minimum, 
with quite infrequent jumps between the wells. As the motion around the 
minimum is approximately harmonic the more interesting part will be the 
transitions between the left and the right potential well, for each 
pair of neighboring particles, going over the hyperbolic saddle at 
$q_{i+1}-q_{i}=0$. Understanding the motion in phase space of a 
high-dimensional system is generally 
very difficult. Some symbolic description of an orbit $\bi{x}(t)$ 
is therefore highly desirable. In IFPU model with high potential barrier the 
choice of a partition of phase space is almost obvious:

For each instant of time $t$ we will be interested only in a single binary bit
of information $a_i \in \{0,1\}$ for each pair of neighboring particles 
$(i,i+1)$, 
namely if the neighbors are in the right or left well, $(q_{i+1}-q_i)>0$ or $< 0$, 
we write $a_i=1$ or $a_i=0$, respectively. 
Binary positions for the whole chain can be compactly encoded in a binary integer 
({\em signature})
\begin{equation}
S=(a_N \ldots a_2 a_1)_2 = \sum_{i=1}^N{a_i 2^{i-1}}.
\label{sig}
\end{equation}
Thus we have defined a partition of $2N$-dimensional phase space cut by 
$N$ hyper-planes $q_{i+1}=q_i$ into $2^N$ cells labeled by signatures 
$S\in \{0,1,\ldots 2^N-1\}$. By a slight abuse of language, we will
use a term {\em signature} $S$ also to refer to a phase space cell denoted by $S$.
If the transitions between various signatures are rare, which is obviously the case
if the potential barrier $\Delta U$ is high, it seems to be meaningful to concentrate
on the (statistical) dynamics of transitions between the signatures. For the periodic boundary conditions the coordinate differences must 
fulfill the constraint $\sum_{i=1}^N{(q_{i+1}-q_i)}\equiv 0$ which in the limit of 
infinitely high potential barrier and even $N$ translates into the condition 
$\sum_{i=1}^N{a_i}=N/2$. For odd $N$ the number of pairs in the left and the right well 
should differ by 1. This presents only unnecessary technical complications and from now on 
we will assume $N$ to be even. If the barrier is finite, signatures with different number 
of pairs in the left and right well are possible to visit. We will call signature $S$ to be of 
order $i$ if $\sum_{i=1}^N{a_i}=N/2 \pm i$. The number of different signatures $M_i$ of 
order $i$ is
\begin{equation}
M_i=\left( {N \atop N/2-i}\right)=\frac{N!}{(N/2-i)!(N/2+i)!}.
\label{M}
\end{equation}  
For sufficiently high barrier the system will spend most of its time in signatures of 
order $0$. We will specifically concentrate on signatures of order 0 and treat signatures
of higher (mostly 1st) order only as 'tunnels' for transitions between different signatures 
of order 0. Figure \ref{Slikaq} shows an example of a transition between two order 0 
signatures via an intermediate short lived signature of order 1. 
First we have to know for which values of parameter $\alpha$ and size $N$
will the description by the signatures of order 0 only be adequate, i.e. 
will the relative time spent in higher order signatures be negligible. 

\subsection{Estimating the fractional volumes of higher order signatures}
\begin{figure}[!ht]
\centerline{\psfig{figure=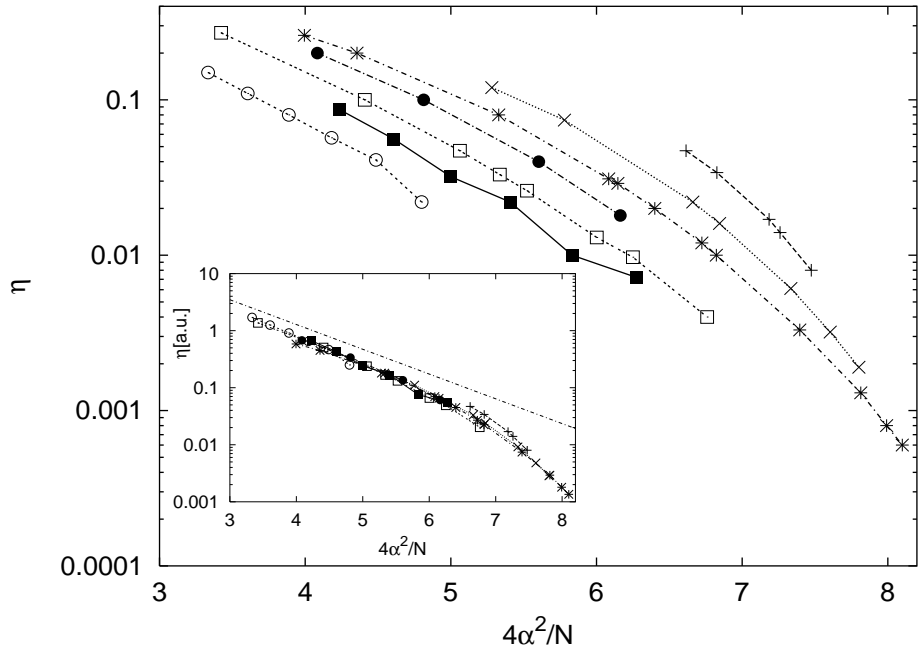,width=130mm,angle=-90}}
\caption{Fraction $\eta=(\sum_{i=1}^\infty t_i)/t_0 \approx t_1/t_0$ of time 
spent in signatures of higher order for $N=6,8,10,12,16,20$ and $30$, from 
right to left, respectively. Lines connect points with the same value of $N$. 
Inset shows by $N$-dependent prefactors rescaled values of $\eta$ and 
exponential function (\ref{nedovol}) (chain straight line).}
\label{Slikanedo}
\end{figure}

We will now make a rough theoretical estimate for the time spent in signatures 
of different orders. For an ergodic system this time is proportional to the measure 
(volume) of the phase space covered with the signatures in question. Let us designate 
by $t_i$ the time spent in signatures of order $i$. We have
\begin{equation}
t_i(E) \propto \int_{{\rm order}\; i}{\frac{dS}{|\nabla H|}}=\frac{d \Gamma_{i}(E)}{d E},
\label{td}
\end{equation}
where the region of integration is the part of the energy surface $H=E$ intersecting with the full set of 
signatures of order $i$ and $\Gamma_i(E)$ is the total volume of the phase space in the
signatures of order $i$ and with energy less than $E$. First, let us make an estimate for 
$\Gamma_0(E)$. Potential around both minima is to the lowest order harmonic, therefore 
we can roughly say $\Gamma_0(E)\approx \Gamma(E)^N$, where $\Gamma(E)$ is the volume 
for one harmonic oscillator, that is $\Gamma(E)=2 \pi E/\sqrt{\alpha}$. For the signatures 
of order $i$ the argument is very similar. In this case there are $2i$ particles more in 
one potential well than in the other. 
Because the trivial condition $\sum_{i=1}^N{(q_{i+1}-q_i)}\equiv 0$ 
must still be satisfied, the equilibrium positions of pairs will not be at 
$\pm \sqrt{\alpha}$ any more but will be shifted for $\Delta q$ in order to keep the 
center of mass of all pairs at 0. This gives for the shift $\Delta q=2i\sqrt{\alpha}/N$ 
and contributes $\varepsilon_i=\alpha \Delta q^2=4i^2 \alpha^2/N^2$ to the energy 
density, provided the cubic part of the potential is negligible, i.e. $2i/N \ll 1$. 
Harmonic approximation for the volume $\Gamma_i(E)$ can therefore still be 
used, but now at energy $E-N\varepsilon_i$. Since we are interested only in the 
dependences for large $N$ we can omit ('cancel') energy derivatives coming from (\ref{td}), so 
that we have
\begin{equation}
\frac{t_i}{t_0} \sim 
\frac{\Gamma(\varepsilon-\varepsilon_i)^N}{\Gamma(\varepsilon)^N}=
\left(1-\frac{\varepsilon_i}{\varepsilon} \right)^N \sim 
\exp{\left(-4i^2 \frac{\alpha^2}{N \varepsilon}\right)}.
\label{nedovol}
\end{equation}
We have fixed $\varepsilon\equiv 1$ so that the relative fraction of signatures of order 1, $t_1/t_0$, falls exponentially in $\alpha^2/N$. We must stress that the exponential dependence is expected only when $4\alpha^2/N^2 \ll 1$ and that the overall prefactor in front of an exponential function could still depend on $N$. This is nicely confirmed by the numerical data in figure \ref{Slikanedo}.
Grouping the signatures by their order is meaningful since members of each group have
the same phase space volume and the same minimal potential energy.
The minimal potential energy of order 0 signatures is $\varepsilon_0=0$, 
of order 1 is $\varepsilon_1=4\alpha^2/N^2$ and of order $i$ is 
$\varepsilon_i \approx i^2 \varepsilon_1$. 
The fraction of time spent in higher order signatures is just a power of relative time spent 
in order 1 signatures, $(t_i/t_0) \approx (t_1/t_0)^{i^2}$ and this can be reduced by increasing $\alpha$.
\par
Of course, for the transitions to be possible at all, the barrier height must be 
smaller than the total energy $E$. This yields the condition 
\begin{equation}
\alpha^2/4 < N
\label{cond}
\end{equation}
which is just the opposite of the condition to keep the $t_1/t_0$ small. 
Nevertheless, we can still reduce the fraction of higher order states to around 
$\exp{(-16)}\sim 10^{-6}$, which is small. But the condition (\ref{cond}) 
can be problematic for small $N$. Indeed, for the smallest nontrivial case of $N=4$ we 
have numerically found that there are no transitions between signatures for $\alpha \ge 2.8$. 
For smaller $\alpha$ we do find transitions, but there the system exhibits non-ergodic behaviour. Condition (\ref{cond}), although 
fulfilled, seems to be too weak and is still keeping the phase space cut
into non-connected parts. 
The smallest system worth studying regarding transitions is therefore $N=6$.
\par
In all numerical work $\alpha$ and $N$ have been chosen so as to keep the fractional
volume of higher order signatures small. We have therefore studied the dynamics on a set of 
$M_0$ signatures of order 0 exhibiting transitions thru short-lived order 1 signatures
as shown in figure \ref{Slikaq}. The most important physical scale that characterizes such
a transport is the average time $\tau$ between subsequent transitions. 

\subsection{Ergodicity and average transition time $\tau$ between signatures of 
order 0}
Though the main body of this paper is concerned with the transport and decay of time-correlations, i.e. with the system's mixing property, we should first mention, that we have also checked the weaker dynamical property of ergodicity directly. We have employed a method of Robnik {\em et al} \cite{Robniketal97} of comparing the rate of visiting of different phase space cells (in out case, order 0 signatures) with the 
rate for a fully random dynamics. The results turned to be fully consistent with (uniformly) ergodic behaviour of our IFPU model in the high-barrier regime discussed above.
\par
\begin{figure}[!ht]
\centerline{\psfig{figure=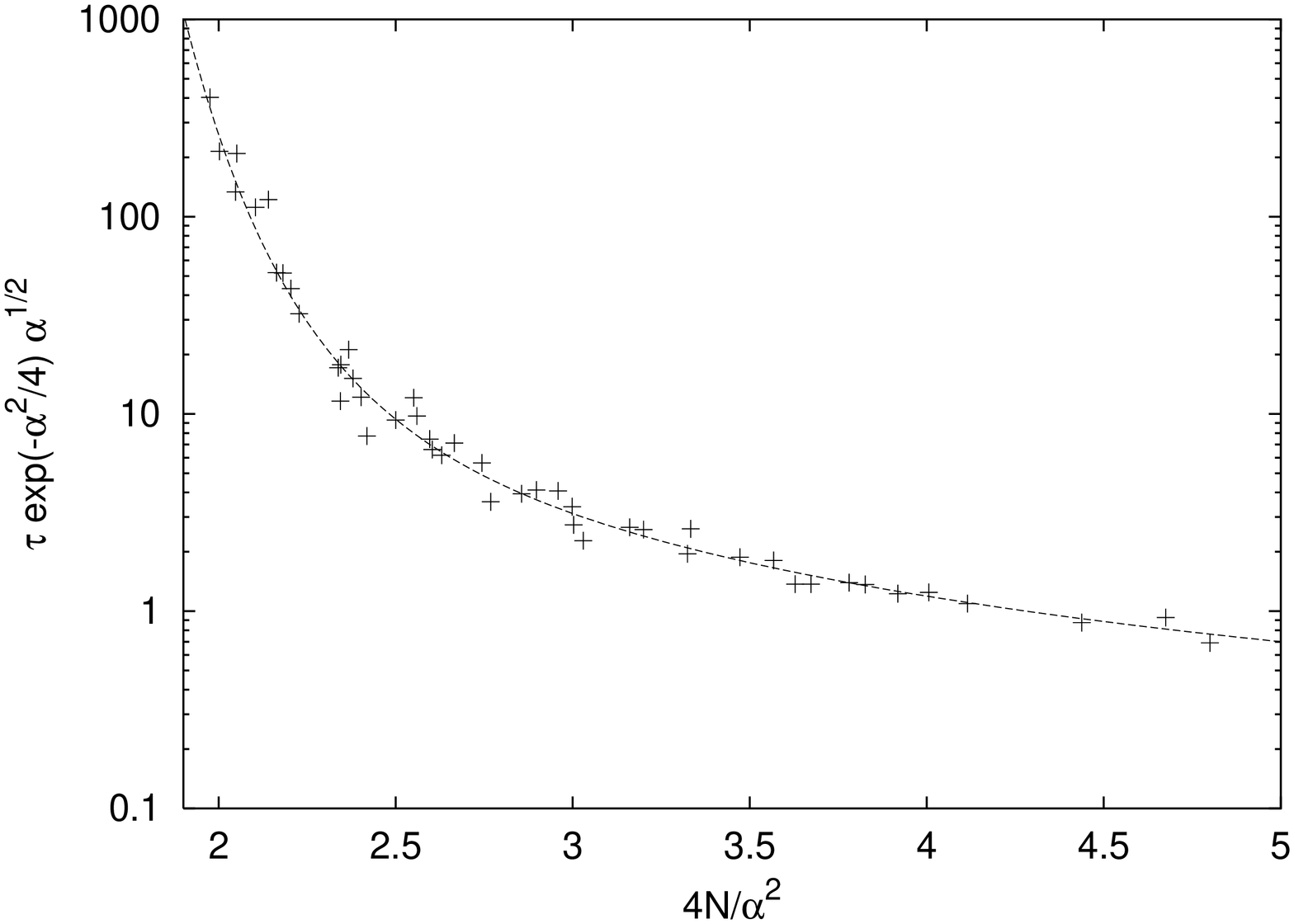,width=100mm,angle=-90}}
\caption{Numerical data for $\tau(N,\alpha)$ divided by 
$\exp{(\alpha^2/4)}/\sqrt{\alpha} $ versus $4N/\alpha^2$. 
The solid curve is only drawn to guide an eye, perhaps suggesting some universal scaling 
function.}
\label{Slikaf}
\end{figure}

Now we define the time scale $\tau$ to be an average time from the point when the
orbit enters a certain order 0 signature $S$ to the point where the same orbit
enters the next order 0 signature $S'$ ($S' \neq S$) concluding the transition from $S$ to $S'$.
The average is taken over many different orbits with microcanonically distributed
initial conditions, or over one very long orbit if we assume ergodicity.
We obtain an approximate functional form for the dependence of $\tau$ on 
$N$ and $\alpha$ by similar arguments as for the relative volume of higher 
order signatures. The probability for a jump will be estimated by the ratio
between the phase space volume of the set of states just before a jump $\Gamma_{\rm t}$ 
and the volume of the set of equilibrium states $\Gamma_{\rm eq}$. The approximate volume
$\Gamma_{\rm eq}(E)$ of the phase space of equilibrium states has been
derived in the previous subsection and is just $\Gamma_{\rm eq}(E) \approx
\Gamma(E)^N=(2\pi/\sqrt{\alpha})^N E^N$. In the state just before the
transitions there should be more than $\Delta U$ of energy in one pair of
particles. This energy will be denoted by $a \alpha^2$, where $a$
is for now some unspecified constant. Certainly $a$ must be greater
than $1/4$, but because our oscillators are coupled we allow for the
possibility that the energy of a pair must actually be bigger than the
barrier height. Phase volume $\Gamma_{\rm t}(E)$ for a state with one pair having 
the energy $a\alpha^2$ and the other $N-1$ pairs residing in the minima is
\begin{eqnarray}
\fl \Gamma_{\rm t}(E)=\int_0^{E-a\alpha^2}{d\varepsilon
\varepsilon^{N-1} \left(\frac{2 \pi}{\sqrt{\alpha}} \right)^{N-1}
(N-1) (\Gamma(E)-\Gamma(\varepsilon))} \nonumber \\
\lo{\approx} \left(\frac{2 \pi}{\sqrt{\alpha}} \right)^N a\alpha^2(E-a\alpha^2)^{N-1}.
\end{eqnarray}
\begin{figure}[!ht]
\centerline{\psfig{figure=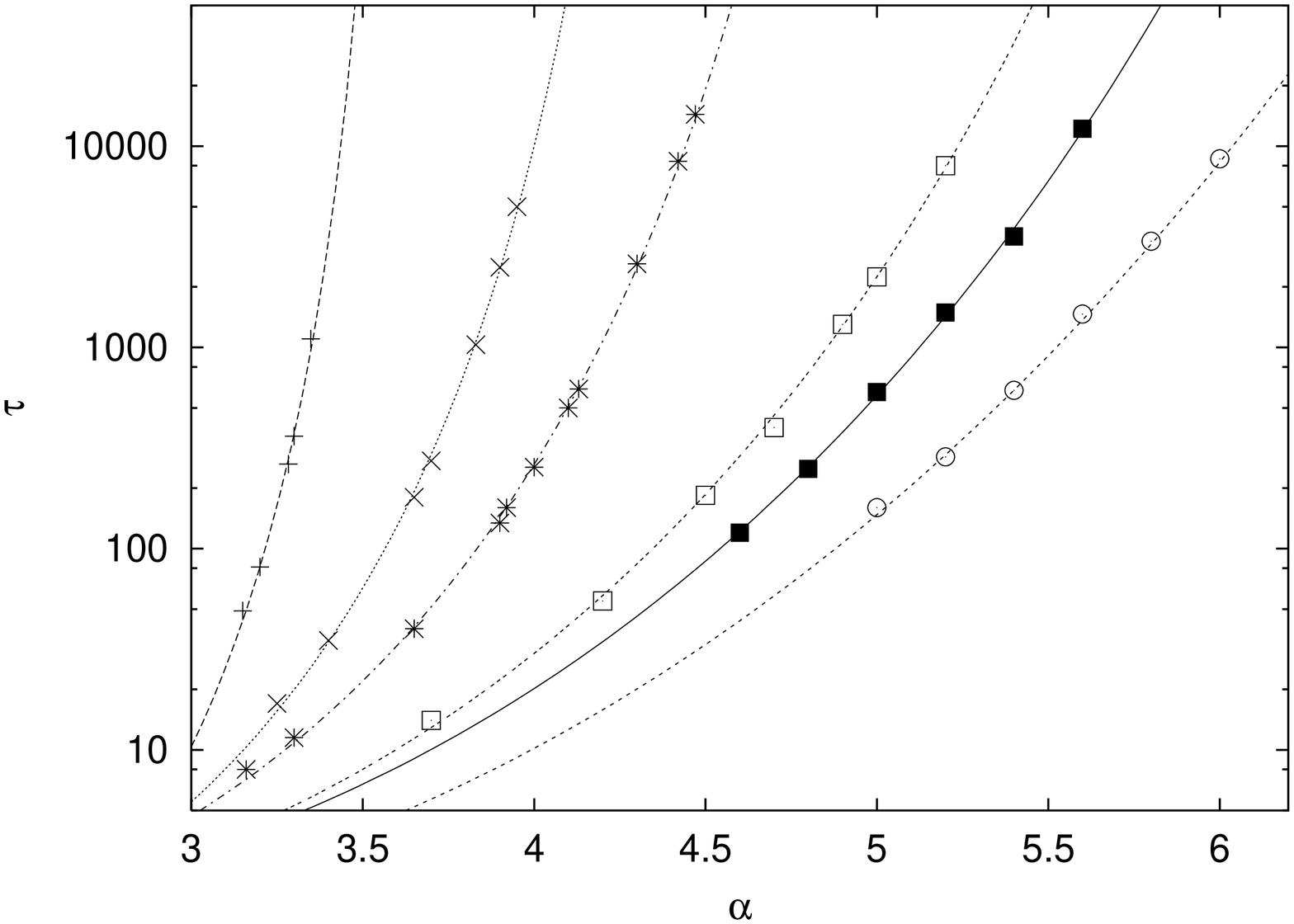,width=130mm,height=90mm,angle=-90}}
\centerline{\psfig{figure=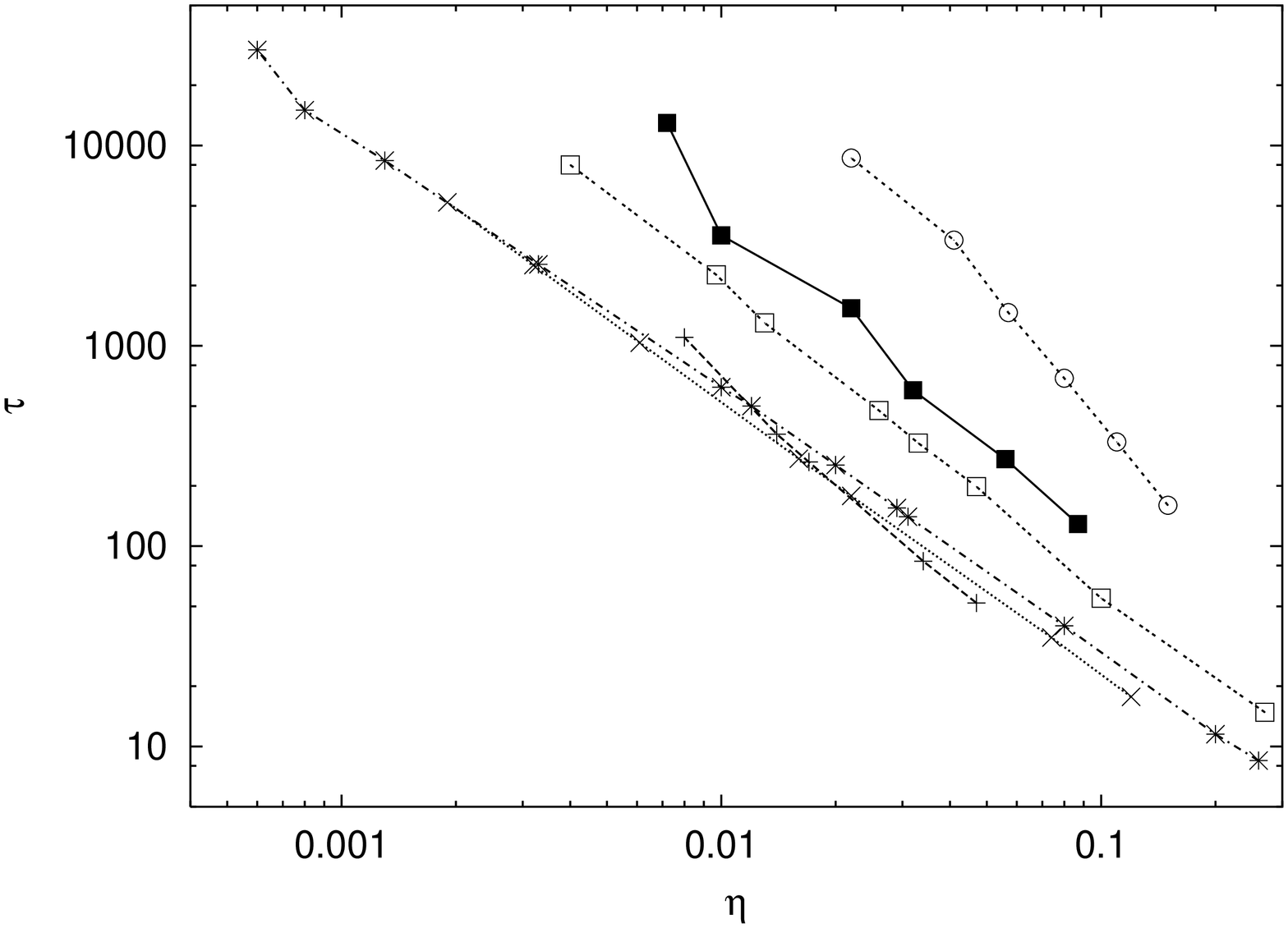,width=130mm,height=90mm,angle=-90}}
\caption{In the top figure experimental values of $\tau$ for $N=6,8,10,16,20$ and $30$ (symbols from left to right) 
and semi-theoretical estimate (with continuous curves) for $\tau$ (\ref{tau}) are drawn. 
Errors in numerical $\tau$ are of 
the same order as the symbol sizes. In the bottom figure we plot the dependence of $\tau$ on 
$\eta=(\sum_{i=1}^\infty t_i)/t_0$. The data and the symbols used are the same as in 
the top figure, while the straight line segments are here merely connecting the points with 
the same value of $N$.}
\label{Slikatau1}
\end{figure}
Probability for a jump is proportional to the quotient of derivatives
$d\Gamma_{\rm t}(E)/d\Gamma_{\rm eq}(E)$. Reciprocal average time
between transitions $1/\tau(N,\alpha)$ is proportional to the above quotient and to the frequency 
of oscillation around potential minimum. Using also $E=N$, we finally get
\begin{equation}
\tau(N,\alpha) \approx \frac{A}{\sqrt{\alpha}} \frac{N}{a \alpha^2}\frac{1}{(1-a\alpha^2/N)^N},
\label{tau}
\end{equation}  
where $A$ is some overall numerical factor. We have determined numerical parameters 
$a$ and $A$ by fitting the numerical data for different lengths $N$ in the range 
$N=6,\ldots,30$. Both parameters do depend on $N$ for small chains, but are, of course,
independent of $\alpha$. The value of $A$ is 
between $0.02$ and $0.05$, whether the $a$ has dependence $a=1/4+{\cal O}(1/N)$. 
In the limit of large chains $a$ has the value $1/4$ predicted by our simple physical 
picture. This picture can also be confirmed by looking at the local energy density just 
before the jump. Indeed, local energy of a pair is exactly $a\alpha^2$ for all chain 
sizes. So, the parameter $a$ is not just some fitting parameter in equation (\ref{tau}) but 
it has a clear physical interpretation. In figure \ref{Slikatau1} we depict the estimate 
of $\tau$ (\ref{tau}) together with the numerical data.
Worth noticing is also an interesting scaling law, although only approximate. Global 
dependence of $\tau(N,\alpha)$ is such that we can write approximately
\begin{equation}
\tau(N,\alpha) \approx \frac{f(4N/\alpha^2)}{\sqrt{\alpha}} \exp{(\alpha^2/4)}, 
\end{equation}
where $f(x)$ is some parameter-free function. The shape of this function can be seen in figure 
\ref{Slikaf}.
\par
\begin{figure}[!ht]
\centerline{
\psfig{figure=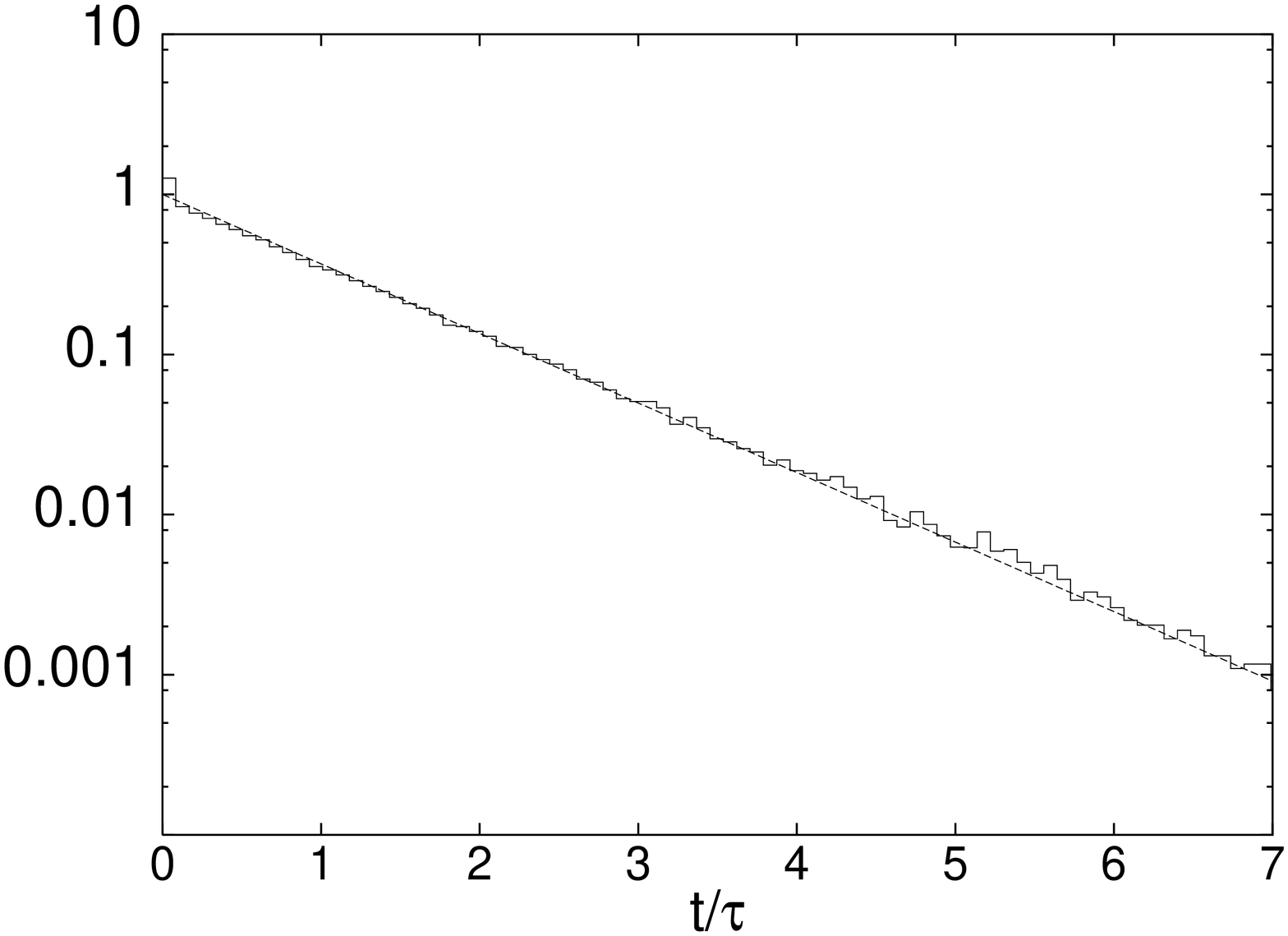,width=77mm,angle=-90}
\psfig{figure=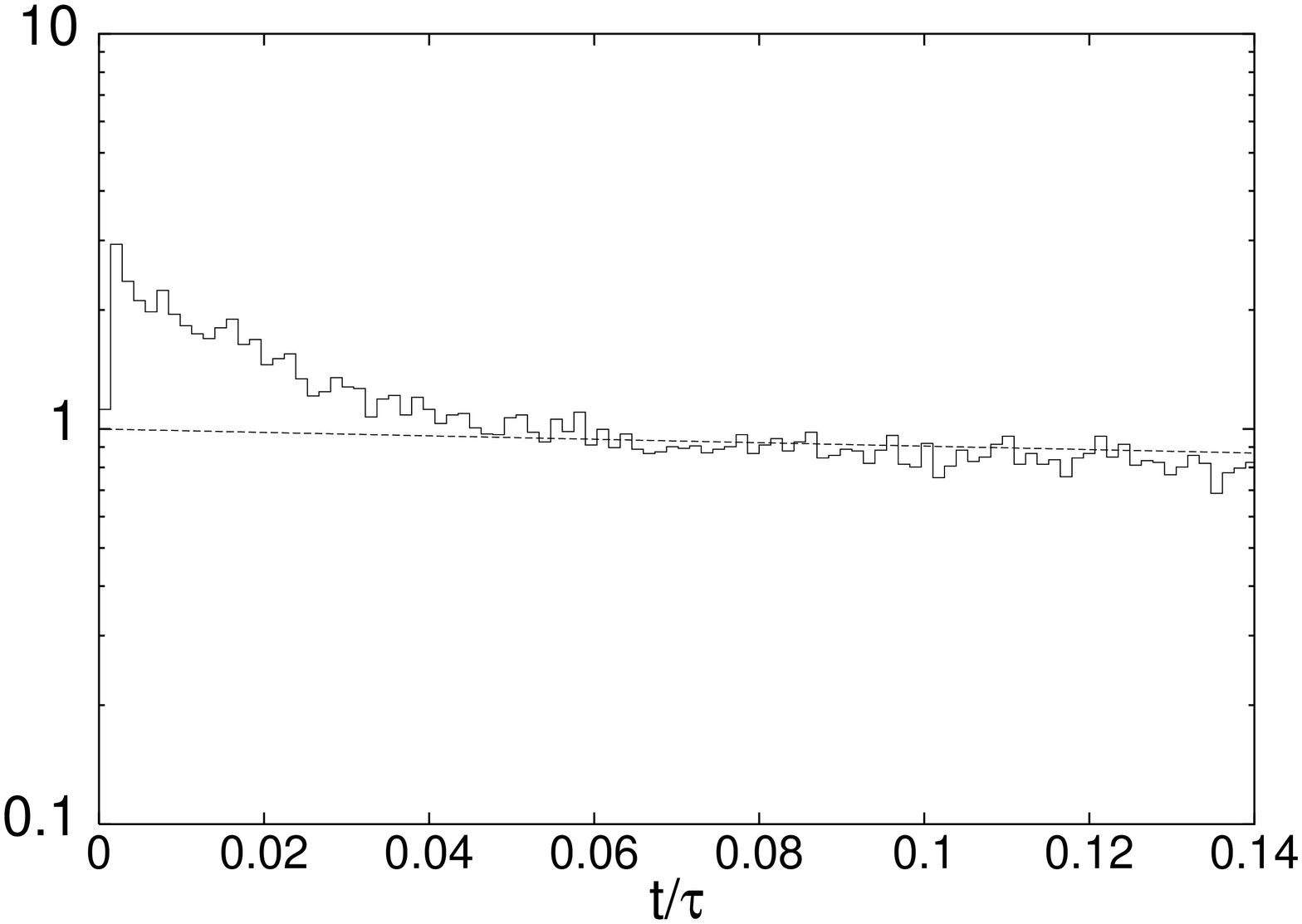,width=77mm,angle=-90}
} 
\caption{Probability distribution of times $t=\tau_n$ (in a relative scale $t/\tau$) 
between two different consecutive order 0 signatures. On the right we plot enlarged 
distribution for small times. All is for the chain of length $N=20$ and $\alpha=5.4$ 
corresponding to $\tau=3500$. Dashed line is an exponential function.}
\label{Slikatau}
\end{figure}

We have also numerically computed statistical distribution of transition times $\tau_n$,
i.e. the time intervals between entering different subsequent signatures of order 0 such that 
$\tau=\< \tau_n\>$, for a fixed $N$ and $\alpha$. This distribution is shown in 
figure \ref{Slikatau} and is clearly seen to be exponential.
There is a discrepancy only for extremely small times. This is a consequence of the 
duration of intermediate unstable order 1 state that occurs between order 0 states 
(see figure \ref{Slikaq}). The fraction of order 1 states was in this case 
$t_1/t_0=0.0013$ which is the same as the time scale of the discrepancy in the figure. 
Our distribution should be a convolution of distributions of $t_0$ and of $t_1$.
But the precise explanation of this discrepancy is the following: we obtain 
too many of ``short'' jumps because of the way we measure
the time when the system has arrived into a 
new order 0 signature. We mark the system as arriving into a new signature exactly at the 
top of the barrier, when the difference $(q_{i+1}-q_i)$ changes the sign and not when it relaxes to the bottom of the well. This 
allows for some `fake transitions' in the regions of phase space where two or more 0 order signatures almost touch with a saddle-like region in between (for instance near coordinate origin $q_i \equiv 0$). There it is possible that the system, which although it gets over the barrier, 
subsequently relaxes into some nearby well because it does not have the 
`right' momentum vector.

\section{Markov process and its transition matrix}
\label{SecMarkov}

We decided to describe dynamics with a sequence of binary signatures instead of a full 
orbit $\bi{x}(t)$ with the hope of some simplification. We are hence looking for some 
simple statistical description of the transitions between signatures. Stochastic system 
is fully described by giving conditional transition probabilities 
$P(S,S',S'',\ldots;t,t',\ldots)$. The simplest kind of a stochastic process is called 
Markov process. For a Markov process the probabilities $P(S,S';t)$ contain entire 
information about the system. We will shortly write a matrix ${\mathbf P}(t)$ 
instead of $P(S,S';t)$. For now let us define $P(S,S';t)$ on the full space of 
$2^N$ signatures although we will later be interested only in transition probabilities
among $M_0$ signatures of order 0. The transition matrix ${\mathbf P}$ must satisfy 
the condition
\begin{equation}
{\mathbf P}(t+t')={\mathbf P}(t) {\mathbf P}(t').
\label{markov}
\end{equation}  
The matrix ${\mathbf P}(t)$ contains therefore some redundant information. We can write 
instead
\begin{equation}
{\mathbf P}(t)=\exp{({\mathbf F}t)},
\label{PizF}
\end{equation}
where for a Markov process the matrix ${\mathbf F}$ is time independent. 
The matrix element $F(S,S')$ is a probability flux from the signature $S$ to the 
signature $S'$, and can be defined by the limit
\begin{equation}
{\mathbf F}=\lim_{t \to 0}{\frac{{\mathbf P}(t)-{\mathbf I}}{t}}.
\label{F}
\end{equation} 
The conservation of probability imposes the condition $\sum_{S'}{F(S,S')}=0$ 
for the matrix elements.  
\par
By a statistical description, using ${\mathbf P}(t)$ instead of $\bi{x}(t)$, 
we lose information about the details within a single cell of a partition. 
In other words, we study the dynamics on the space of functions that are 
constant over one phase space cell. This can be formalized by introducing 
characteristic functions 
$B_S(\bi{x})$ over signatures $S$
\begin{equation}
B_S(\bi{x})=
\left\{
\begin{array}{ll}
1 & \bi{x} \in S \\ 
0 & \hbox{otherwise}
\end{array}
\right. .
\label{karakterS}
\end{equation}
Characteristic functions $B_S$ span the space of all observables 
which are piece-wise constant over the signatures. 
Every piece-wise constant observable $W(\bi{x})$ can be expanded over the base
functions $B_S$ as ${\rm W}(\bi{x})=\sum_S{w_S B_S(\bi{x})}$ with the expansion 
coefficients $w_S$ given by
\begin{equation}
w_S(t)=\frac{\langle B_S {\rm W}(t) \rangle_{\rm E}}{\langle B_S
\rangle_{\rm E}},
\label{razvojr}
\end{equation}
where the brackets $\langle \rangle_{\rm E}$ denote a microcanonical 
phase space average over the energy surface $H=E$. 
The matrix elements of a transition matrix ${\mathbf P}$ are nothing but 
the correlation functions between the characteristic functions
\begin{equation}
P(S,S';t)=\langle B_{S'}(t) B_S \rangle_{\rm E}/\langle B_S
\rangle_{\rm E}.
\label{PzB}
\end{equation}
As the transition matrix propagates the probabilities between the signatures we can write 
the time dependent vector ${\bi w}(t)=(w_S(t),w_{S'}(t),\ldots)$ as
\begin{equation}
{\bi w}(t)={\mathbf P}(t) {\bi w}(0)=\exp{({\mathbf F}t)} {\bi w}(0).
\label{evolr}
\end{equation}
Similarly, the autocorrelation function of the observable $W$ is 
\begin{equation}
\langle {\rm W}(t) {\rm W}(0) \rangle_{\rm E}={\bi w}^T \exp{({\mathbf F}t)} {\bi w}.
\label{avtow}
\end{equation}
We can see that the behaviour of correlation functions is determined by the spectrum of 
the (Markov) propagator $\exp{({\mathbf F}t)}$.
\par
From now on we assume that the fraction of higher order signatures is very small
$\eta\approx t_1/t_0 \ll 1$ and we restrict the matrices ${\mathbf P}(t)$ and
${\mathbf F}$ on the $M_0$-dimensional subspace of signatures of order 0, 
which essentially contain
all non-vanishing matrix elements. 
Numerically we calculated the transition matrix ${\mathbf P}(t)$ from one very long orbit 
as an average
\begin{equation}
P(S,S';t)=\frac{\langle \delta_{S',S(t'+t)} \delta_{S,S(t')}
\rangle_{\rm{t'}}}{\langle \delta_{S,S(t')} \rangle_{\rm{t'}}},
\label{Pt}
\end{equation}
where the brackets $\langle \rangle_{\rm{t'}}$ denote a time average over the
orbit and $\delta_{S,S(t)}$ has value 1 if the orbit is in a signature $S$ at 
time $t$ and 0 otherwise. For small $t$ the average can be performed, and by using 
expression (\ref{F}), yields for $F(S,S')$,  $(S \neq S')$ 
\begin{equation}
F(S,S')=\frac{n(S,S')}{t_S},
\label{Fss}
\end{equation}
where $n(S,S')$ is the number of direct transitions between the
order 0 signatures $S$ and $S'$ and 
$t_S$ is the total time spent in the signature $S$. Here a small comment regarding 
higher order signatures is in order. We saw in figure \ref{Slikaq} that between each
order 0 states there is a short intermediate order 1 state of duration
$\sim \tau t_1/t_0$. So there are no direct transitions between order 0 states 
and all fluxes $F(S,S')$ (\ref{Fss}) among $M_0$ order 0 signatures are strictly zero. 
As we have decided to study only (indirect) transitions between order 0 states 
(for the parameter values where the fraction of higher order states is negligible) 
we must somehow circumvent this difficulty. One solution is to calculate the 
derivative of the transition matrix ${\mathbf P}(t)$ in equation (\ref{F}) not at time 
$t=0$ but at the time $t$ with $\tau t_1/t_0 \ll t \ll \tau $. In this case the number 
$n(S,S')$ in (\ref{Fss}) is the number of jumps between order 0 states $S$ and 
$S'$ with one short intermediate order 1 state. Still more simple and convenient solution is to replace the orbit $\bi{x}(t)$ by a sequence of times $t_i$ when the orbit hits the boundary of a cell of order 0 coming from outside. Orbit $\bi{x}(t)$ is therefore replaced by a sequence of pairs $\ldots,(S_i,t_i),(S_{i+1},t_{i+1}),\ldots$ which tell you that at 
time $t_i$ orbit $\bi{x}(t)$ came to the order 0 signature $S_i$, then until 
time $t_{i+1}$ was in signature $S_i$ or any higher order signature and at time 
$t_{i+1}$ arrived into the next order 0 signature $S_{i+1}$ ($S_{i+1} \neq S_i$), etc.
Then we define $n(S,S')$ as the number of subsequent pairs $(S,S')$ in the 
sequence $\{S_i\}$, and $t_S = \sum_i^{S_i=S}(t_{i+1}-t_i)$, hence the flux matrix is calculated as
\begin{equation}
F(S,S') = \frac{\sum_i^{(S,S')=(S_i,S_{i+1})} 1}{\sum_i^{S_i=S}(t_{i+1}-t_i)}.
\end{equation}
This definition is used in the numerical calculation of this paper.    

\subsection{Tests for the Markov process}
Before describing the process of transitions as Markovian we must check if this 
is at all permissible. Checking the rigorous conditions for the partition to signatures to 
be Markovian, see e.g. \cite{gaspard}, is a considerable mathematical problem,
at least to our judgement. On the other hand, heuristic arguments and numerical results 
support the idea, that for sufficiently large transition times $\tau$ the process will 
indeed be Markovian. 
The system has positive 
Lyapunov exponents.\footnote{We have numerically checked that the 
distribution of Lyapunov exponents is approximately linear, 
as is typically the case for sufficiently chaotic systems \cite{Livietal87}.} 
Vaguely this means that the system quickly ``forgets'' its history. 
If the average time between the jumps $\tau$ is longer than the Lyapunov time, we can 
expect transitions to be Markovian. However, the most straightforward test is to check 
explicitly the composition formula (\ref{markov}).
\par
\begin{figure}[!ht]
\centerline{\psfig{figure=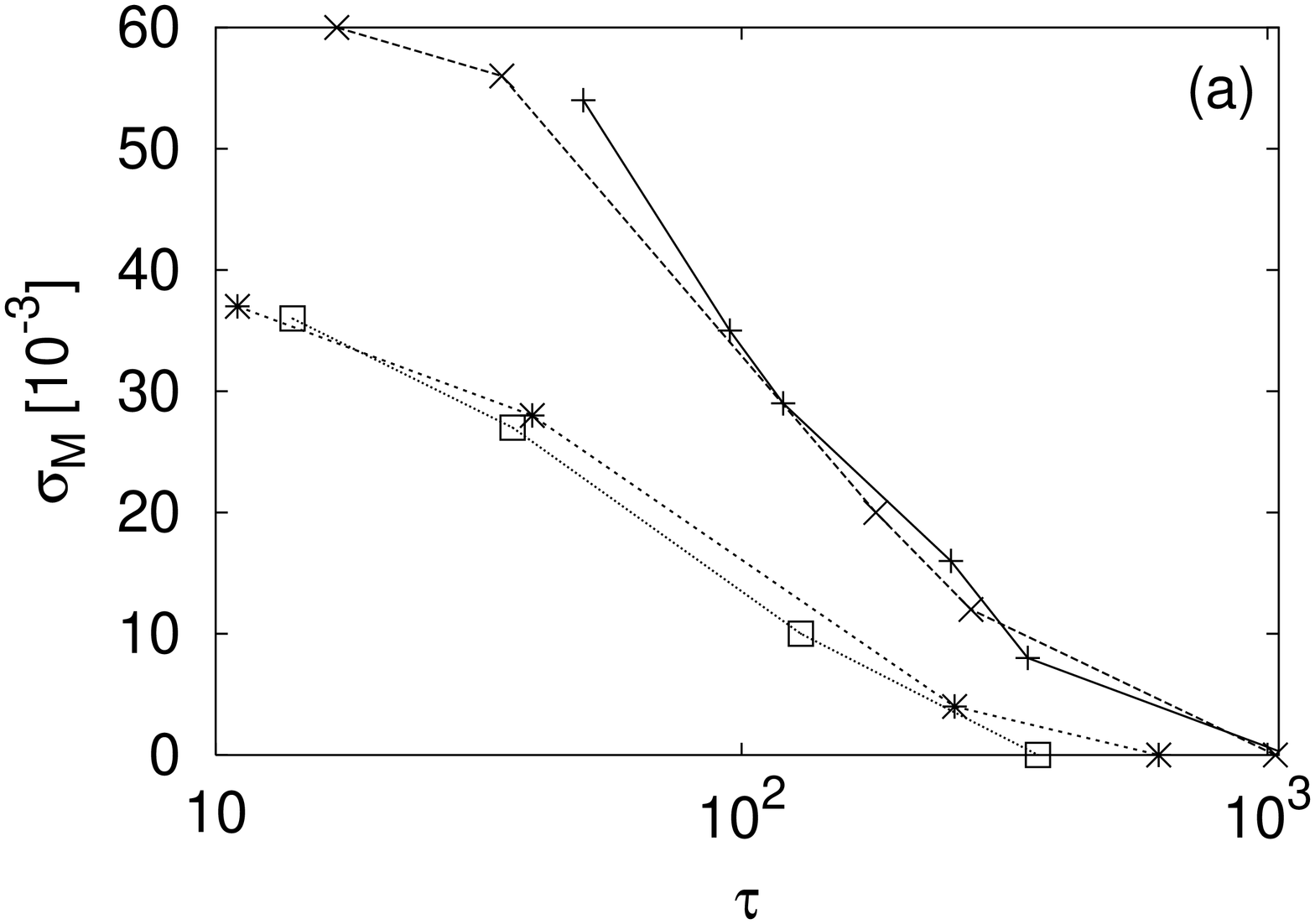,width=78mm,angle=-90}
\hskip -7 mm
\psfig{figure=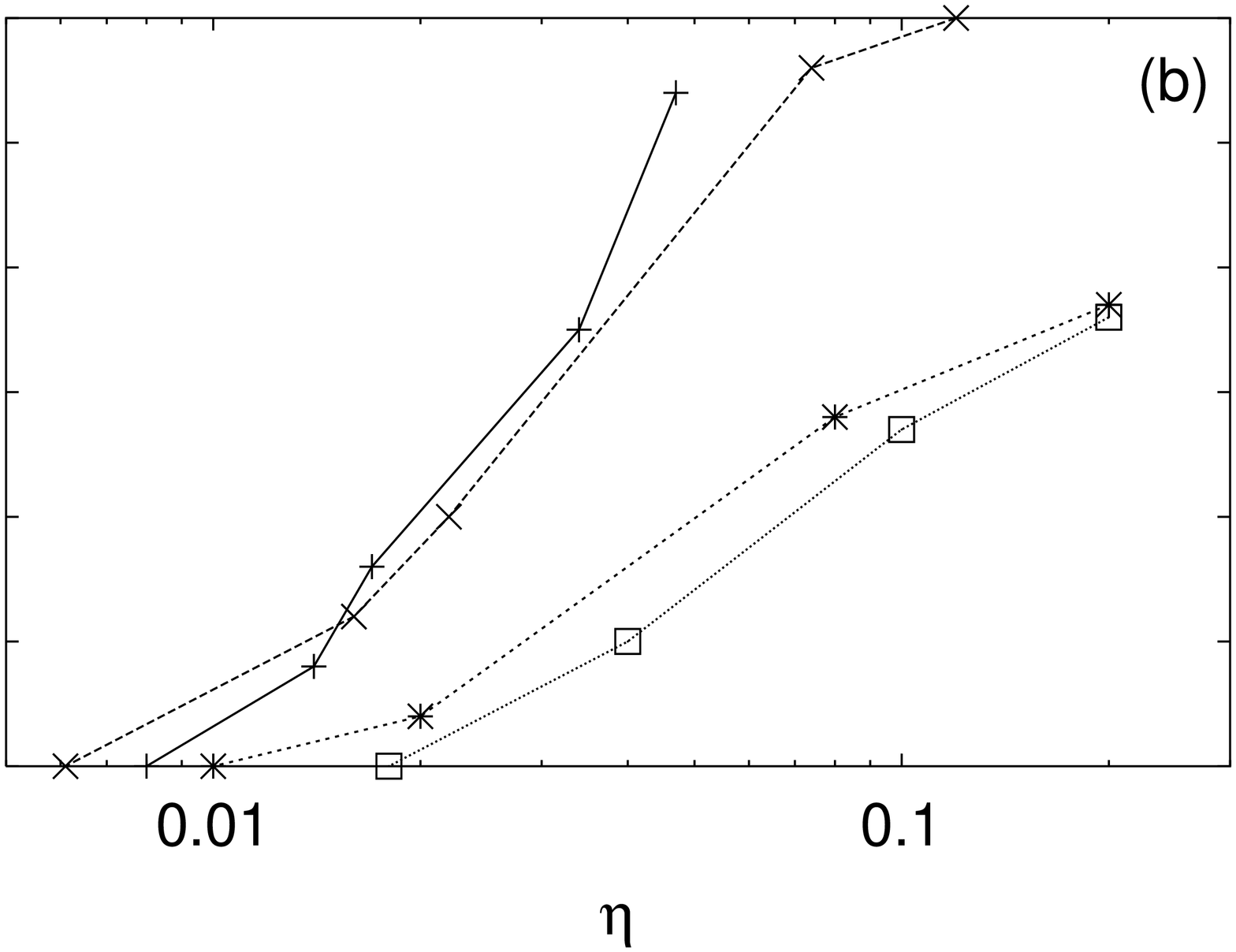,width=78mm,angle=-90}}
\caption{Dependence of $\sigma_{\rm M}$ on $\tau$ in figure (a) and on 
$\eta=(\sum_{i=1}^\infty t_i)/t_0$ in figure (b). Pluses, crosses, stars and squares are for the $N=6$,8,10 and 12, respectively.}
\label{SlikaMark}
\end{figure}

Condition ${\mathbf P}(2t)={\mathbf P}^2(t)$ is trivially fulfilled for small and for 
large times $t$, $t\ll\tau$ and $t\gg\tau$, respectively.  
We checked it for time $t=\tau$ by calculating the quantity
\begin{equation}
\sigma=\frac{||{\mathbf P}(2\tau)-{\mathbf P}^2(\tau) ||_{\rm E}}{||{\mathbf
P}(2\tau)||_{\rm E}},
\label{sigma}
\end{equation} 
where $||{\mathbf A}||_{\rm E}=(\sum_{i,j} |A_{i,j}|^2)^{1/2}$ is the Euclidean norm of a 
matrix ${\mathbf A}$ (all results have also been
re-checked by using the spectral norm giving almost identical results). 
As the matrix ${\mathbf P}(t)$ will
be calculated numerically from one very long but finite orbit it will have
statistical error $\sigma_{\rm T}$ due to the finite number of simulated
transitions. In addition, $\sigma$ will also have a contribution
$\sigma_{\rm M}$ from a systematic error because the process may not be
precisely Markovian. We have assumed $\sigma^2=\sigma_{\rm
M}^2+\sigma_{\rm T}^2$ and made an estimate for the statistical error
$\sigma_{\rm T}$ as a square root of a number of average transitions per matrix element $n$. It is 
\begin{equation}
\sigma_{\rm T}=c \frac{1}{\sqrt{n}}=c \frac{N}{2}\sqrt{\frac{M_0}{N_0}},
\label{sigmat}
\end{equation}
where $N_0$ is the total number of all transitions between the signatures of 
order 0 and $c$ is some unspecified numerical constant which has been
determined from the numerical data. Its value turned out to be $c\approx 0.7$. Results
for the dependence of a systematic error $\sigma_{\rm M}$ are shown in figure 
\ref{SlikaMark}.
The error $\sigma_{\rm M}(\tau)$ goes to zero with increasing $\tau$, as expected. 
Even more interesting is the fact that the error also seems to decrease with increasing 
$N$ at constant $\tau$. This is interesting as it means that with increasing $N$ we 
do not have to keep the fraction of higher order states small in order for the process 
to stay Markovian. 
\footnote{In such a case one could bring higher order signatures back and describe a 
complete hierarchy of transitions between signatures of various orders as a Markov 
process on $2^N$ dimensional space.}
\par
\begin{figure}[!ht]
\centerline{\psfig{figure=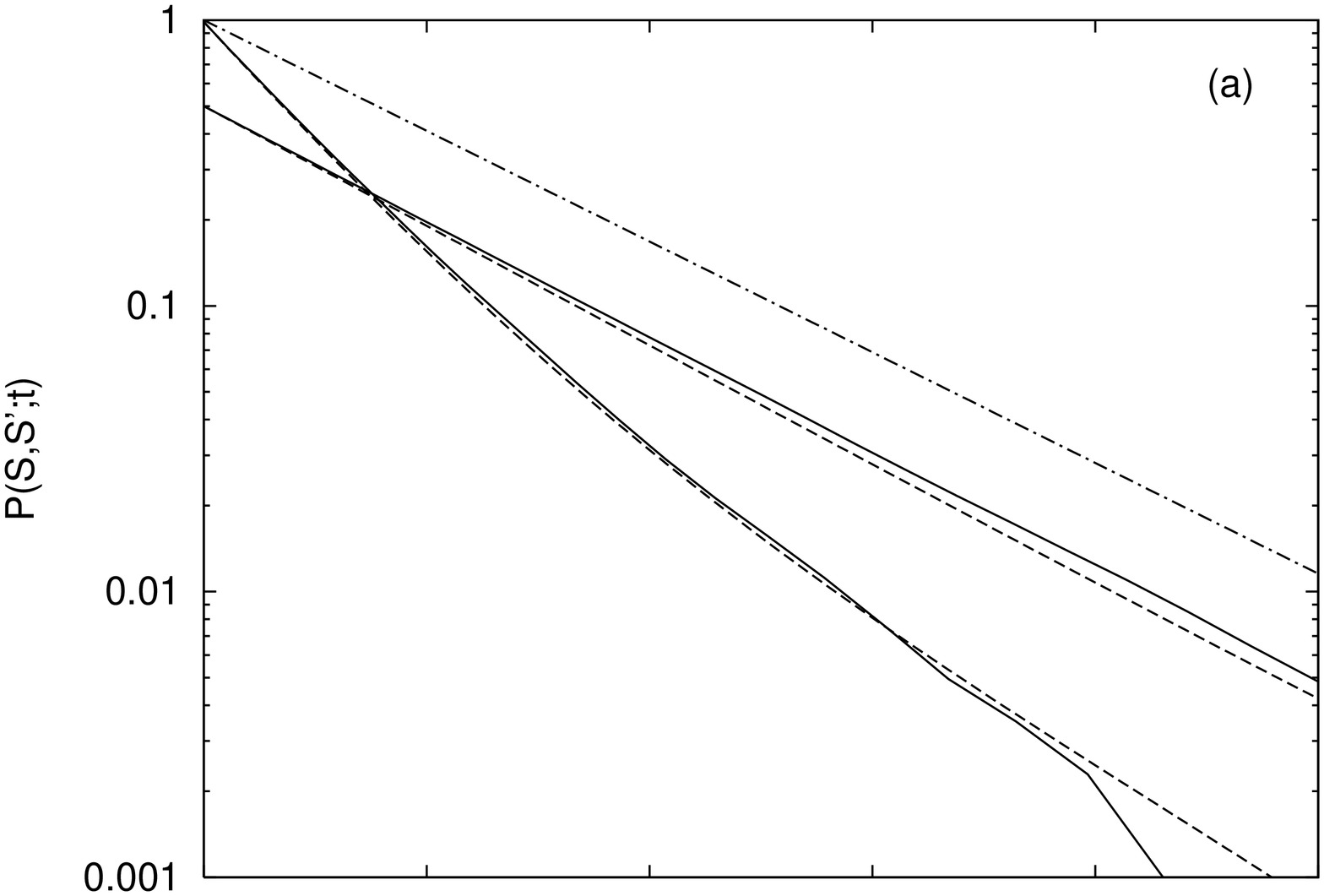,width=130mm,height=70mm,angle=-90}}
\centerline{\psfig{figure=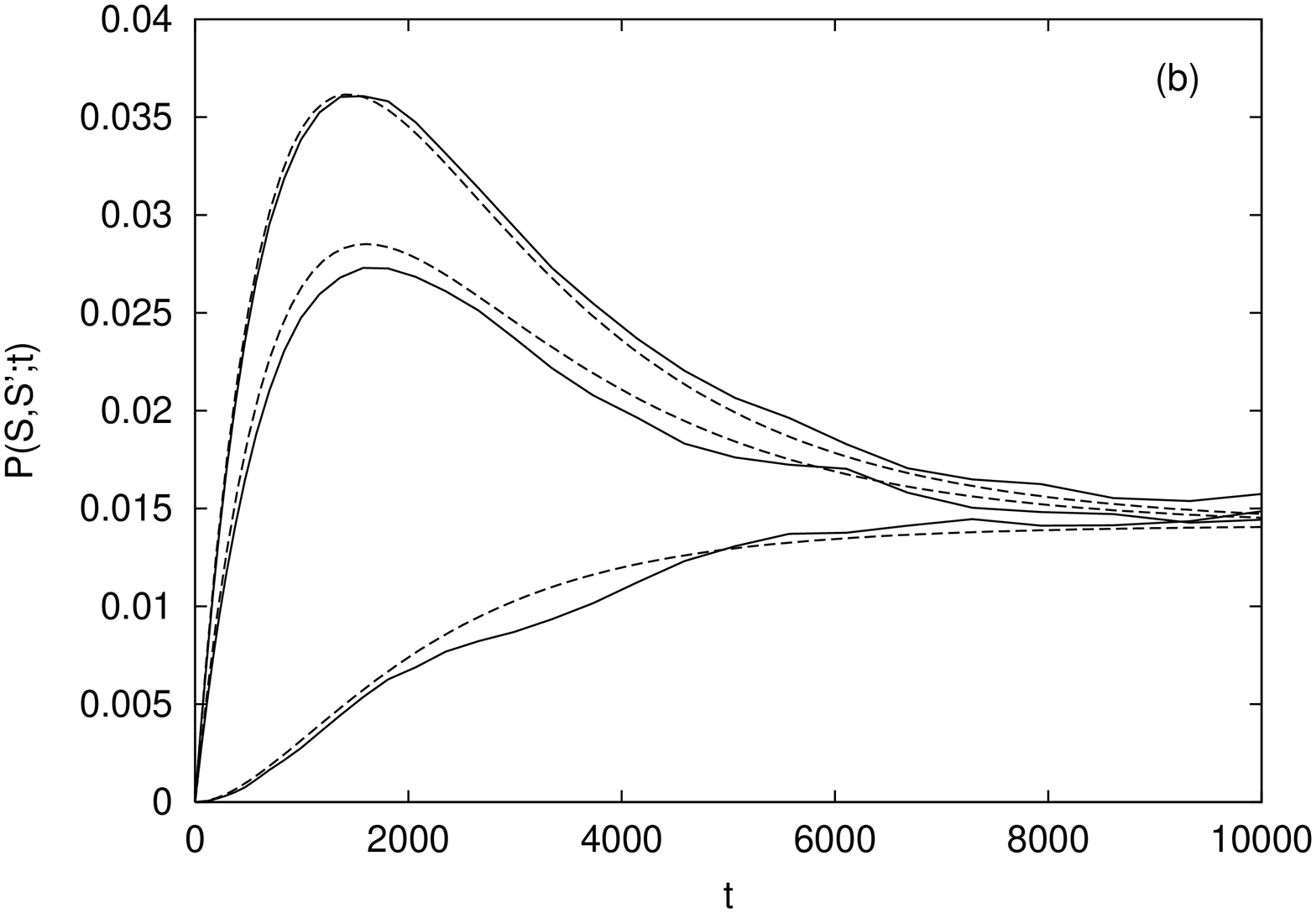,width=130mm,height=75mm,angle=-90}}
\caption{Comparison of the correlation functions calculated from the 
propagator $\exp{(\mathbf F t)}$ (dashed curves) and from directly determined 
transition matrix $P(S,S';t)$ (full curves). In figure (a) are autocorrelation functions 
$S=S'$ and in figure (b) cross-correlation functions $S\neq S'$. 
Bottom pair of curves in figure (a) is for the signature $S=54$, 
the middle pair for the one particle state $\ket{s1}$ and the top 
curve is a simple exponential decay with the biggest nonzero eigenvalue of the 
matrix ${\mathbf F}$. From all correlation 
functions we have subtracted equilibrium value which is 
$P(S,S;t\to \infty)=1/M_0$ and $1/2$ for the state $\ket{s1}$. 
In figure (b) are correlation functions between the signature $S=54$ and 
the signature $S'=58$, $S'=116$, and $S'=135$, from top to bottom, respectively. 
The chain length and the parameter were $N=8$, $\alpha=3.83$, $\tau \sim 1000$. 
}
\label{Slikan81}
\end{figure}

Whether a given system can be described as a Markov process can be checked also by 
calculating the transition matrix ${\mathbf P}(t)$ and comparing it with the exponential
formula (\ref{PizF}) as a function of time. ${\mathbf P}(t)$ can be calculated 
numerically directly from definition, for instance by using the formula (\ref{Pt}). 
This definition would have a meaning also if the transitions were not Markovian.  
We then compare the correlation functions between the characteristic 
functions (\ref{PzB}) and the matrix elements of the propagator 
$\bra{S}\exp{({\mathbf F}t)}\ket{S'}$. These calculations should
agree only if the flux matrix ${\mathbf F}(t)=\log{{\mathbf P}(t)}/t$ were time 
independent and this would signal that the transitions are described by a 
Markov process. Numerical results for this test are shown in figure \ref{Slikan81}.
Correlation functions calculated in both ways have been computed for
the characteristic function $B_S\equiv\ket{S}$ on the signature $S=54$ and for the 
{\em macroscopic observable} with the vector
$w_S=S \hbox{ mod } 2$, i.e. characteristic function on a union of half of all signatures
with the first bit $a_1=1$, denoted by $\ket{s1}$. It can be seen from the figure that 
the autocorrelation function of the observable $\ket{s1}$ begins to fall as 
$\exp{(-\lambda_1 t)}$, where $\lambda_1$ is the biggest nontrivial eigenvalue of 
${\mathbf F}$, very early. 
This is a consequence of relatively large overlap between the macroscopic
state $\ket{s1}$ and 
the eigenstate $\ket{v1}$ corresponding to the eigenvalue $\lambda_1$. 
We will show in subsection \ref{dot} that $|\braket{s1}{v_1}|^2 \sim 1/N$ while 
$|\braket{S}{v_1}|^2 \sim 1/M_0$, so $|\braket{s1}{v_1}|^2 \gg |\braket{S}{v_1}|^2$ 
(for $N\gg 1$). 

\section{The form and the spectral properties of the flux matrix}
\label{SecF}

Something can be said immediately about the spectrum of ${\mathbf F}$. 
Conservation of probability implies all eigenvalues $\lambda_i$ to be smaller or 
equal to zero. Further, because the system is believed to be ergodic, there should be 
a non-degenerate constant eigenvector with 
the corresponding eigenvalue $\lambda_0=0$. This is all the 
consequence of the Perron-Frobenius form of the matrix 
${\mathbf P}$. For finite $N$ the spectrum 
of the flux matrix is discrete and the correlation functions will fall asymptotically as 
$\exp{(-\lambda_1 t)}$ where $\lambda_1$ is the largest nonzero eigenvalue. 
An interesting question is what happens in the thermodynamic limit $N \to \infty$. 
If there is a spectral 
gap we will have an exponential decay, otherwise some anomalous behaviour can occur.

\subsection{Numerical calculation of the flux matrix}

Non-vanishing matrix elements of ${\mathbf F}$ are expected only between the signatures 
that differ just by two bits provided that $t_1/t_0 \ll 1$. In other words, 
two bits $a_i=1$ and $a_j=0$ only switch 
their position. Such a transition will be called a (bit) 
jump of length $d$, where $d=|j-i|$ is 
the distance between the two bits involved. Example of a such transition has been shown in 
figure \ref{Slikaq}. We have numerically calculated complete matrices ${\mathbf F}$ for 
sizes $N=6,8,10$ and $12$ and different $\tau$ (or $\alpha$), in order to check the above 
hypothesis and to obtain some knowledge about the sizes and the structure of matrix 
elements. The only matrix elements which are really non-vanishing are 
those of the bit jumps, and what is more, the value of the matrix element 
depends only on the distance $d$ of the jump 
and not on the particular signatures involved. This practically means that in addition to 
the diagonal elements we have only $N/2$ different matrix elements in the flux matrix 
${\mathbf F}$. This number must be compared to the total number of matrix elements 
$M_0^2$ which grows exponentially with $N$. We have checked this also for the matrices 
${\mathbf F}$ of sizes up to $N=30$ but in this case we have compared only the 
jumps of length $d$ with different number of bits 1 on the sites between 
the two jump sites. We confirmed that the probability fluxes of jumps depend only on 
the length $d$. It is therefore meaningful to 
introduce dimensionless coefficients 
$c_d$ which are ratios between the matrix elements of a jump of length $d$ and 
a jump of length $d=1$. Therefore $c_1\equiv 1$, by definition, and the 
rest of $N/2-1$ 
coefficients $c_d$ together with the average transition time $\tau$ is all 
that we need in order to specify the flux matrix ${\mathbf F}$ completely. 
We can write
\begin{equation}
{\mathbf F}=\frac{1}{\tau}\frac{2N-2}{N^2(c_{N/2}/2+\sum_{d=1}^{N/2-1}{c_d})} {\mathbf C},
\label{c}
\end{equation}
where the matrix ${\mathbf C}$ is just a matrix involving the coefficients 
$c_d$ only. 
Numerical values of $c_d$ for different $N$ and $\tau$ are listed in table \ref{c_d}.
It can be seen that for not too big $\tau$ (e.g. $\tau=11$ and $N=10$) the coefficients $c_d$  
decrease monotonically with the distance $d$. For bigger $\tau$ (e.g. $\tau=621$ and $N=10$), this dependence ceases to 
be monotonic but becomes slightly well-shaped with a minimum at around $d\approx
N/4$. 
\begin{table}[t]
\hspace{4mm}
\begin{minipage}[t]{53mm}
\vspace{0pt}
\begin{tabular}{|r|c|c|c|}
\hline
\multicolumn{4}{|c|}{N=6} \\ \hline
$\tau$ & $\eta [10^{-2}] $ & $c_2$ & $c_3$ \\ \hline \hline
50 & 4.7 & 0.71 & 0.64 \\
360& 1.4 & 0.72 & 0.66 \\
1100&0.8 & 0.73 & 0.66 \\ \hline
\end{tabular}
\end{minipage}
\hspace{5mm}
\begin{minipage}[t]{86mm}
\vspace{0pt}
\begin{tabular}{|r|c|c|c|c|c|c|}
\hline
\multicolumn{7}{|c|}{N=12} \\ \hline
$\tau$ & $\eta [10^{-2}] $ & $c_2$ & $c_3$ & $c_4$ & $c_5$ & $c_6$ \\ \hline \hline
37 & 10 & 0.79 & 0.68 & 0.63 & 0.62 & 0.63 \\
130& 4.0 & 0.81 & 0.73 & 0.71 & 0.76 & 0.79 \\
366& 1.8 & 0.83 & 0.76 & 0.76 & 0.85 & 0.92 \\ \hline
\end{tabular}
\end{minipage}

\vspace{2mm}
\hspace{4mm}
\begin{minipage}[t]{64mm}
\vspace{0pt}
\begin{tabular}{|r|c|c|c|c|}
\hline
\multicolumn{5}{|c|}{N=8} \\ \hline
$\tau$ & $\eta [10^{-2}] $ & $c_2$ & $c_3$ & $c_4 $ \\ \hline \hline
17 & 12 & 0.72 & 0.61 & 0.57 \\
180& 2.2 & 0.76 & 0.67 & 0.65 \\
1030&0.6 & 0.76 & 0.71 & 0.71 \\ \hline
\end{tabular}
\end{minipage}
\hspace{6mm}
\begin{minipage}[t]{74mm}
\vspace{0pt}
\begin{tabular}{|r|c|c|c|c|c|}
\hline
\multicolumn{6}{|c|}{N=10} \\ \hline
$\tau$ & $\eta [10^{-2}] $ & $c_2$ & $c_3$ & $c_4 $ & $c_5$ \\ \hline \hline
11 & 20 & 0.74 & 0.61 & 0.54 & 0.52 \\
250& 2.0 & 0.80 & 0.74 & 0.76 & 0.78 \\
621&1.0 & 0.82 & 0.76 & 0.82 & 0.86 \\ \hline
\end{tabular}
\end{minipage}
\caption{Dependence of $c_d$ on distance $d$ for different $\tau$ and $N$.}
\label{c_d}
\end{table}
There also seems to be a general trend that with increasing $N$, at a constant 
fraction $\eta \approx t_1/t_0$, the coefficients $c_d$ all approach 1. This can be
explained by the following heuristic argument. To keep the fraction $t_1/t_0$ of
higher order states constant with increasing $N$ we must increase the
barrier height $\Delta U$. As a consequence, the transition time $\tau$ 
increases and correspondingly also
the average time $\tau t_1/t_0$ spent in the non-equilibrium intermediate state
of order 1 in between the states of order 0. At a constant $t_1/t_0$
the transition time $\tau$ depends approximately exponentially on the chain
length $N$ (see figure \ref{Slikatau1}). On the other hand, the relaxation time for the 
non-equilibrium energy distribution at the moment of transition grows perhaps only
linearly with $N$ (inversely proportional to the sound speed), but certainly
slower than exponentially. For
sufficiently large $N$ the relaxation time will therefore be smaller
than the time spent in the intermediate order 1 cell. Initial locally
non-equilibrium distribution at the beginning of a jump will therefore
relax before the transition to the new order 0 state, which will therefore all be equally probable. Because of
this we can assume that the coefficients $c_d$ will not depend on $d$ in the thermodynamic
limit. This hypothesis still needs further verification as the limit of high 
$N$ could not be tested due to the fast growth of the dimension $M_0$ with $N$. Despite of that, 
the numerical results are consistent and point in the right direction 
(see table \ref{c_d}). 
We will show now, that our flux matrix has a specially appealing form which can be
described using a formal connection to a certain Hamiltonian of a quantum spin-$1/2$ chain.
Based on this we will also deduce the essential spectral properties of the flux matrix.

\subsection{Correspondence between the Markov process and the quantum spin chain}
We can make the announced correspondence by connecting the eigenvalue problem for
the flux matrix $\mathbf F$ with the eigenvalue problem for the quantum Hamilton
operator $\rm \hat H$ over $M_0$-dimensional Hilbert space with the same matrix 
elements as $\mathbf F$. This correspondence would generally be of no particular 
use since the Hamiltonian $\rm H$ does not represent any simple quantum system. 
In our example of IFPU chain this is not the case, since $\mathbf F$ has a particularly
nice and simple form, with non-vanishing elements only between the signatures connected
by a bit jump where signatures are sequences of binary bits, 0 and 1.
This should immediately remind us of Heisenberg chains of quantum spin-$1/2$ 
particles. Let us write the Hamilton function for one-dimensional 
ferromagnetic Heisenberg spin chain of length $N$, 
with Pauli variables $\vec{\sigma}_j=(\sigma_j^{\rm x},\sigma_j^{\rm y},\sigma_j^{\rm z})$ 
and coupling constants $J(d)$
\begin{equation}
\fl {\rm \hat H}=-\frac{1}{2}\sum_{j,d=1}^N{ J(d) \vec{\sigma}_j \cdot
\vec{\sigma}_{j+d}}=-\frac{1}{2}\sum_{j,d=1}^N{J(d)\left\{
\sigma_j^{\rm z}
\sigma_{j+d}^{\rm z}+\frac{1}{2}\left( \sigma_j^+ \sigma_{j+d}^- + \sigma_j^- \sigma_{j+d}^+ \right) \right\} },
\label{Heis}
\end{equation}
where periodic boundary condition is assumed $\mathbf{\sigma}_{N+j} \equiv
\mathbf{\sigma}_j$ and $J(d) \equiv J(N-d),J(0) \equiv 0$. Operators
$\sigma_j^\pm=\sigma_j^{\rm x} \pm {\rm i} \sigma_j^{\rm y}$ are the standard 
raising and lowering operators. Quantum state of the spin chain, an eigenstate of $\sigma_j^{\rm z}$, 
will be denoted by the signature $\ket{S}$, with the obvious interpretation. 
Bit $a_j=1$ or $0$ denotes spin $j$ up or down,
respectively. Now we can calculate the matrix elements of the Hamilton
operator (\ref{Heis}). For the off diagonal elements we get
\begin{equation}
\fl
\bra{S} {\rm \hat H} \ket{S'}=\left\{ \begin{array}{cl}
 -2J(d) & \hbox{if there is a jump of length $d$ connecting $S$ and $S'$} \\
 0 & \hbox{otherwise}
\end{array} \right. .
\label{izven}
\end{equation}
Diagonal elements are 
\begin{equation}
\bra{S} {\rm \hat H} \ket{S}= -J(N/2) \left[ N/2-2 s_{N/2}(S) \right]-\sum_{d=1}^{N/2-1}{ J(d) \left[ N-2 s_d(S) \right]},
\label{diag}
\end{equation}
where $s_d(S)$ is a number of different signatures that can be
reached from the signature $S$ with a jump of length $d$. For any
signature $S$ we can write an identity
\begin{equation}
\sum_{d=1}^{N/2}{s_d(S)}=\left( \frac{N}{2} \right)^2.
\label{sumsos}
\end{equation}
By denoting $s=M_0 N^2/(2N-2)$, the following equality is also valid
\begin{equation}
\sum_{i=1}^{M_0}{s_d(S_i)}=\left\{ 
\begin{array}{cl}
 s & d=1,\ldots N/2-1 \\
 s/2 & d=N/2 \\
\end{array} \right. .
\label{s}
\end{equation}
\par
The ground state for the ferromagnetic Heisenberg Hamiltonian (\ref{Heis}) 
can be found immediately for any $J(d) > 0$. We have to keep in mind that the 
Hilbert space is in our case spanned just by $M_0$ signatures of order 0 and not by all
possible signature states as is usual for the quantum spin chains. Order
$i$ of the signature is simply an eigenvalue of the $\rm S_z$ component of the
total spin, namely ${\rm {S_z}}=\sum_j{
\sigma_j^{\rm z}/2}=i$. But the operator $\rm S_z$ commutes with the
Hamilton function and the Hilbert space is therefore a direct sum of
subspaces labelled by eigenvalues of $\rm S_z$. 
We choose $M_0$ dimensional Hilbert subspace with 
${\rm S_z}(=i)=0$. Ground state denoted by $\ket{0}$ is
\begin{equation}
\ket{0}=\frac{1}{\sqrt{M_0}} \sum_{i=1}^{M_0}{\ket{S_i}},
\label{zero}
\end{equation}
with the energy
\begin{equation}
E_0 \equiv \bra{0} {\rm \hat H} \ket{0}
=-N \left( \sum_{d=1}^{N/2-1}{J(d)}+\frac{1}{2}J(N/2) \right).
\label{E0}
\end{equation}  
\par
Now that we have the matrix elements of $\rm H$ and the ground state, we can see that if we
prescribe $J(d)$ to equal $c_d$ multiplied by the constant
prefactor in front of the matrix $\mathbf C$ in equation for $\mathbf F$ (\ref{c}), 
we can formally write our flux matrix in terms of the Heisenberg spin Hamiltonian (\ref{Heis}) 
\begin{equation}
{\mathbf F}=-{\mathbf H}+E_0{\mathbf I}.
\label{H'}
\end{equation}
Transition probabilities $\rm{P}(S,S';t)$ can now be written as
\begin{equation}
{\mathbf P}(t)=\exp{({\mathbf F}t)}=A(t) \exp{(-\beta{\mathbf H} )},
\label{kvantkanon}
\end{equation}
where we wrote $\beta=t$ and $A(t)=\exp{(E_0 t)}$. Expression for
$\mathbf P$ has the same form as the density operator for the quantum
canonical distribution at temperature $1/t$. Time dependence of
$\mathbf P$ is therefore the same as the dependence of the canonical
distribution on cooling. Understanding time dynamics of the IFPU model is 
equivalent to the cooling of the ferromagnetic Heisenberg spin
chain or its imaginary time dynamics. In the limit $t \to \infty$ the matrix elements of $\mathbf P$ go
towards the constant value $1/M_0$ and in the corresponding 
Heisenberg spin chain any non-equilibrium distribution relaxes to the
ground state $\ket{0}$.
\par
\begin{figure}[ht]
\centerline{\psfig{figure=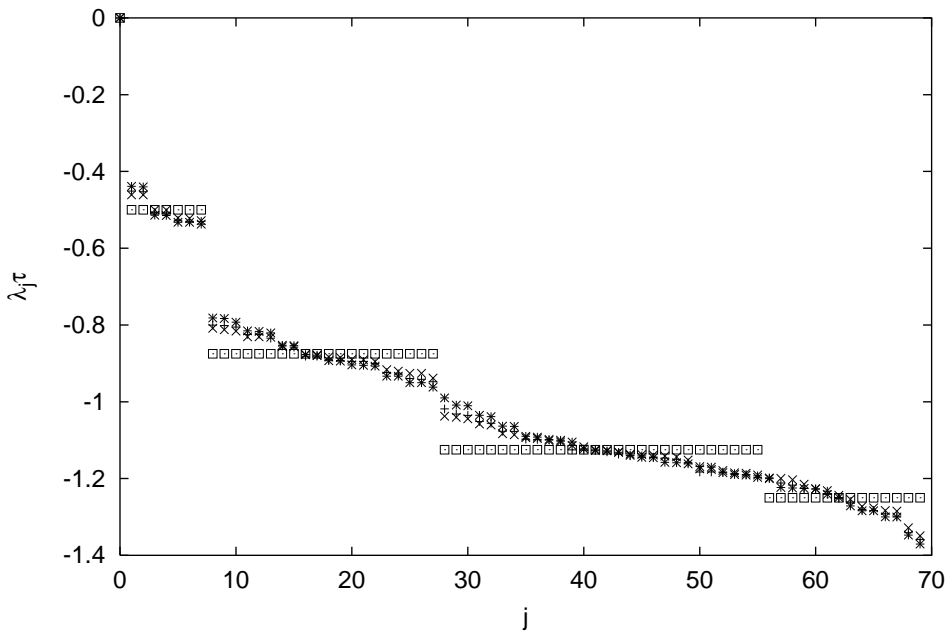,width=130mm,angle=-90}}
\caption{Spectrum of $\mathbf F$ for $N=8$ at three different
$\tau=17$, $180$ and $1030$ with stars, pluses and crosses, respectively. 
With boxes is plotted the spectrum in the case $c_d \equiv 1$ 
(\ref{lambdai}) for comparison.}
\label{Slikalam8}
\end{figure}
\begin{figure}[!ht]
\centerline{\psfig{figure=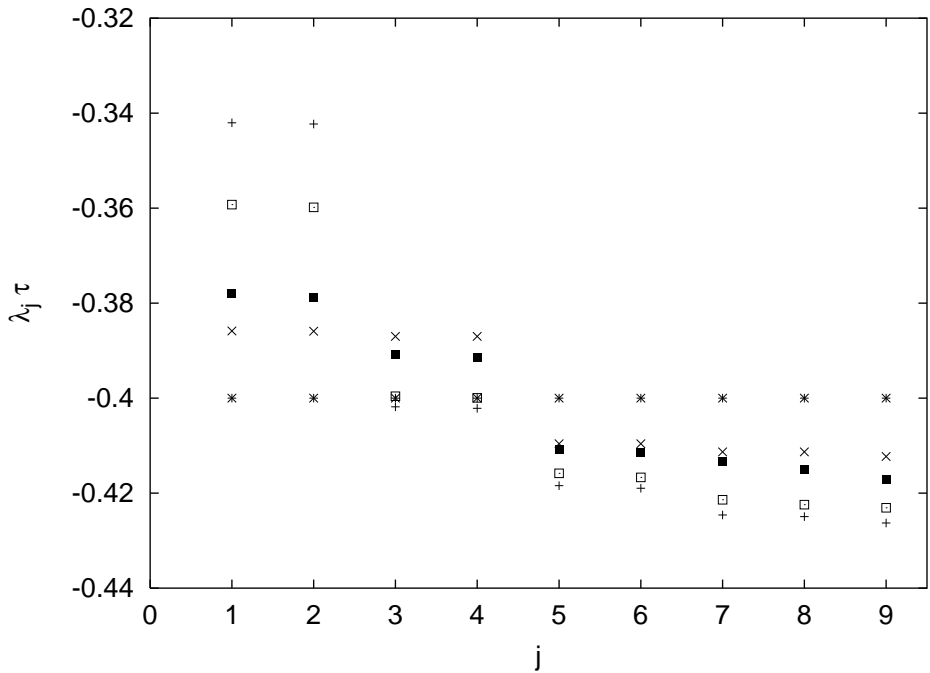,width=130mm,angle=-90,rheight=90mm}}
\caption{Enlarged first multiplet  
for $N=10$ and $\tau=11$,40,250 and 620 with pluses, squares, full squares and crosses, 
respectively. Referential degenerate set of first $N-1$ eigenvalues at $c_d\equiv 1$ is 
plotted with stars}
\label{Slikalam102}
\end{figure}

For higher lying states we numerically solved the eigenvalue problem for
the flux matrix at various $\tau$ and sizes up to $N=12$. One
such example of a numerical spectrum is shown in figure \ref{Slikalam8}. 
We have said that with increasing $N$, at constant $t_1/t_0$, we expect coefficients
$c_d$ to approach a constant value $c_d=1$ independent of $d$. This case 
corresponds to the simplest Heisenberg spin chain with uniform coupling. Apart from the
constant factor the Hamilton function (\ref{Heis}) is in this case
${\rm H}=-2 {\mathbf S}^2+3N/2$ and the matrix $\mathbf F$ reads
\begin{equation}
{\mathbf F}=\frac{2}{N^2 \tau}\left( 2 {\mathbf S}^2-\frac{N(N+2)}{2}{\mathbf I} \right),
\label{S}
\end{equation}
where 
${\rm \bf S}^2=1/4(\vec{\sigma}_1+\ldots+\vec{\sigma}_N)^2$. Eigenvalues of
$\mathbf S^2$ are $S(S+1)$, where $S$ is the quantum number of the total spin with
values from $S=0,\ldots,N/2$. The Hilbert space is composed
of order 0 states with $\rm S_z=0$. If we denote by $\lambda_j$
the eigenvalues of operator (\ref{S}) and by $n(j)$ the corresponding
multiplices we have
\begin{eqnarray}
\lambda_j &=&-\frac{4j(N+1-j)}{N^2 \tau}  \nonumber \\
n(j)&=&\frac{(N+1-2j) N!}{(N+1-j)! (j)!}  \qquad j=0,\ldots,N/2. 
\label{lambdai}
\end{eqnarray}
Particularly interesting are the first two eigenvalues $\lambda_0$ and
$\lambda_1$. The biggest, non degenerate eigenvalue $\lambda_0=0$ belongs
to the ground state $\ket{0}$. The largest nontrivial
eigenvalue is $\lambda_1=-4/N\tau$ with the multiplicity $n(1)=N-1$. In the
thermodynamic limit and keeping $\tau$ constant, it goes to zero as
$1/N$. The correlation functions therefore decay {\em slower than exponential} 
in the thermodynamic limit. From the figures
\ref{Slikalam8} and \ref{Slikalam102} it can also be deduced that
with increasing $\tau$ the spectrum is indeed approaching the spectrum
for the $c_d\equiv 1$ (\ref{lambdai}), as predicted. What is more, the
eigenvalue $\lambda_1$ seems to be monotonically approaching the limiting case
($c_d\equiv 1$) from above. We can therefore conclude by observation that the 
correlation functions of observables which can be spanned by $\ket{S}$ decay 
slower than $\exp{(-4t/N\tau)}$. 
\par
\label{dot}
Now that we expect the spectrum to be close to the one for $c_d \equiv 1$ we
can explain the decay of the correlation functions for {\em one particle state}
$\ket{s1}$ and the {\em one signature state} $\ket{S}$ in figure
\ref{Slikan81}. State vector for $\ket{S}$ has only one component
different from 0. Eigenvector $\ket{v1}$ corresponding to the
eigenvalue $\lambda_1$ has on the other hand all components
approximately of the same order, i.e. $1/\sqrt{M_0}$. Square of the scalar product
is therefore $|\braket{S}{v_1}|^2 \sim 1/M_0$. One particle state
$\ket{s1}$ is proportional to the $\ket{s1} \sim \sigma_1^+ \sigma_1^-
\ket{0}$. Because we act on the state $\ket{0}$ with the eigenvalue
$S=N/2$ with the operator $\sigma_1^+ \sigma_1^-$, we can decompose the
state $\ket{s1}$ as a linear combination of states with the eigenvalue $S=N/2$ and
$S=N/2-1$. State $\ket{s1}$ can therefore be written as a sum of $N$
states, one of which is also $\ket{v1}$. If we assume all the
expansion coefficients to be of the same order we immediately obtain
$|\braket{s1}{v_1}|^2 \sim 1/N$.

\section{Discussion and conclusion}

We divide the phase space of a $N$-body Hamiltonian, namely the inverted FPU 
model, into $2^N$ cells that are uniquely tagged by a binary number, 
called a signature (\ref{sig}). 
Signatures are {\em ordered} according to the absolute difference in the 
number of particle pairs/bits in the left/0 and right/1 potential well. 
For a sufficiently high potential barrier an approximate 
Markovian description on space spanned by  
order 0 signatures is possible. The accuracy of the 
Markovian property has been checked numerically to be 
increasing with increasing chain length $N$ at a constant fraction of higher 
order states. In the flux matrix $\mathbf{F}$ we have
non-vanishing matrix elements only between the signatures connected by 
an exchange of a pair of bits. This enables for the
formal correspondence between the IFPU and ferromagnetic Heisenberg quantum 
spin-$1/2$ chains. Understanding time dependence of a Markov transition matrix 
is equivalent to cooling or imaginary time dynamics of a quantum spin chain. 
If we increase the chain length $N$ at a fixed fraction of higher order
states, the transition rates are expected to become independent of a 
jump length $d$. Such trend is confirmed by a numerically calculated matrices 
$\mathbf{F}$. In this limit exact eigenvalues of a flux matrix can be obtained
explicitly. The largest nontrivial eigenvalue in this case is 
$\lambda_1=-4/N \tau$ and goes to $0$ in the thermodynamic limit. There is also a numerical evidence that the largest 
nontrivial eigenvalue in a general case (for a non-constant transition rate) 
is strictly larger than $-4/N \tau$. Therefore, the correlation functions of 
the observables which can be spanned by functions $\ket{S}$ decay asymptotically
slower than $\exp{(-4t/N\tau)}$. 
\par
Unfortunately these results do not imply directly the behaviour of
more general correlation functions, such as current-current correlation
which is needed in order to understand the transport properties.
We have made some numerical calculations of current-current time correlation 
functions of IFPU chain for different sizes $N$ and different parameters. As expected, 
the decay of correlations for observables that vary on time scales smaller 
than $\tau$ (e.g. current) is faster than the decay of correlation functions 
spanned by the piece-wise constant basis $\ket{S}$. Transition from the 
algebraic (anomalous conductivity) to the exponential (normal conductivity) 
decay of current autocorrelation function occurs rather abruptly at 
$\alpha \approx 3$ (for $N \ge 20$).
\par
Finally, though we suggested several interesting properties of the
above model, we must admit that most of our conclusions are based on
numerical evidence. Therefore, we believe that it should be a challenging 
and not impossible future task to try to provide more rigorous justifications
(and perhaps proofs) of our results. Establishing rigorous asymptotic Markovian 
property would enable one to systematically code and enumerate all the many-body 
(unstable, hyperbolic) periodic orbits and to use them explicitly in a classical or 
semi-classical {\em trace formulae}, for example to calculate the transport coefficients 
directly. We feel that IFPU chain may become a useful toy model of a 
{\em chaotic field theory} \cite{Cvitanovic99}.

\section*{Acknowledgement}
Financial support by the Ministry of Science and Technology of Slovenia
is gratefully acknowledged.

\end{document}